\documentclass[english,useAMS,usenatbib]{mn2e} 
\usepackage{lmodern}

\usepackage[T1]{fontenc}
\usepackage[latin9]{inputenc}
\usepackage{color}
\usepackage[usenames,dvipsnames]{xcolor}
\usepackage{amsmath}
\usepackage{amssymb}
\usepackage{graphicx}
\usepackage{esint}
\usepackage{placeins}
\usepackage[authoryear]{natbib}

\usepackage{hyperref}
\definecolor{darkgreen}{rgb}{0.1,0.6,0.7}
\hypersetup{colorlinks = true,urlcolor=blue,citecolor=Blue}

\makeatletter
\usepackage{lmodern}\usepackage{epstopdf}

\newcommand{\half}{\frac{1}{2}}

\newcommand{\boX}{\ensuremath{\boldsymbol{X}}}
\newcommand{\bomu}{\ensuremath{\boldsymbol{\mu}}}
\newcommand{\boSig}{\ensuremath{{\sf{\Sigma}}}}
\newcommand{\sfS}{\ensuremath{{\sf{S}}}}
\newcommand{\sfA}{\ensuremath{{\sf{A}}}}
\newcommand{\sfF}{\ensuremath{{\sf{F}}}}
\newcommand{\mathd}{\ensuremath{\mathrm{d}}}

\newcommand{\both}{\ensuremath{\boldsymbol{\theta}}}
\newcommand{\obsboX}{\ensuremath{\boX_{\mathrm{o}}}}

\newcommand{\boXbar}{\ensuremath{\bar{\boX}}}

\newcommand{\fom}{\ensuremath{{\rm FoM}}}
\newcommand{\var}{\ensuremath{\mathcal{V}}}

\title[Lost information due to estimated covariance matrices]{Quantifying lost information due to covariance matrix estimation in parameter inference }

\author[E. Sellentin \& A.F. Heavens]{Elena Sellentin, Alan F. Heavens\\
Imperial Centre for Inference and Cosmology (ICIC), Department of Physics, Imperial College, Blackett Laboratory,\\ Prince Consort Road, London SW7 2AZ, U.K.
}

\makeatother

\usepackage{babel}
\begin{document}

\date{Accepted 17th October 2016. Received 17th October 2016; in original form 6th September 2016}

\maketitle
\pagerange{\pageref{firstpage}--\pageref{lastpage}} \pubyear{2016}

\label{firstpage} 
\begin{abstract}
Parameter inference with an estimated covariance matrix systematically loses information due to the remaining uncertainty of the covariance matrix. Here, we quantify this loss of precision and develop a framework to hypothetically restore it, which allows to judge how far away a given analysis is from the ideal case of a known covariance matrix. We point out that it is insufficient to estimate this loss by debiasing a Fisher matrix as previously done, due to a fundamental inequality that describes how biases arise in non-linear functions. We therefore develop direct estimators for parameter credibility contours and the figure of merit, finding that significantly fewer simulations than previously thought are sufficient to reach satisfactory precisions. We apply our results to DES Science Verification weak lensing data, detecting a 10\% loss of information that increases their credibility contours. No significant loss of information is found for KiDS.
For a Euclid-like survey, with about 10 nuisance parameters we find that 2900 simulations are sufficient to limit the systematically lost information to 1\%, with an additional uncertainty of about 2\%. Without any nuisance parameters 1900 simulations are sufficient to only lose 1\% of information. We further derive estimators for all quantities needed for forecasting with estimated covariance matrices. Our formalism allows to determine the sweetspot between running sophisticated simulations to reduce the number of nuisance parameters, and running as many fast simulations as possible.
\end{abstract}


\section{Introduction}
For surveys of the cosmic large-scale structure as for example in \citet{DES, Heymans,Euclid,Synergies}, cosmology currently finds itself in the situation of having difficulties reliably describing measurement uncertainties via a covariance matrix. Whereas analytical approximations for the covariance matrix exist and have been exploited e.g.\ in KiDS \citep{KiDS}, covariance matrices are usually estimated from numerical simulations with the hope that these model the specifics of a survey and the non-linear structure growth better than analytical approximations \citep{HD, Blot,Sato}. 

Uncertainty in the covariance matrix inevitably leads to loss of information, and a modified likelihood function \citep{SH15}.  Here we investigate further the loss of information, to quantify the expected gains or losses obtainable by increasing or decreasing the number of simulated datasets.

\begin{figure}
\includegraphics[width=0.45\textwidth]{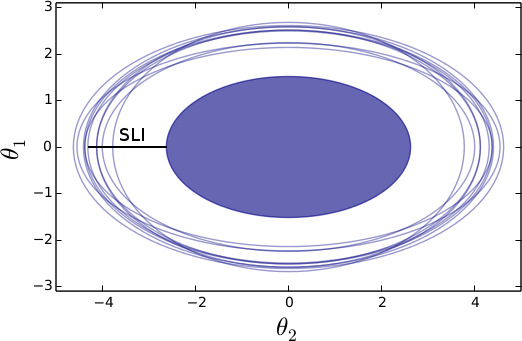} 
\caption{Toy model, estimating linear parameters from a $t$-distribution with multiple Wishart-sampled covariance matrices (open contours), or from a Gaussian distribution with the true covariance matrix (solid contour). The gap between the solid contour and the open contours illustrates the systematically lost information (SLI) due to the uncertain covariance matrix $\sfS$. This gap diminishes when increasing the number of simulations $N_s$.
}
\label{ToyModel}
\end{figure}

We begin with an unbiased sample estimator for the covariance matrix. If $N_s$ independent simulations each yield a synthetic data vector $\boX_i$, with the sample average being $\boXbar = \frac{1}{N_s}\sum_{i = 1}^{N_s} \boX_i$,
then an unbiased estimator of the true but unknown covariance matrix $\boSig$ is
\begin{equation}
 \sfS = \frac{1}{N_s-1} \sum_{i = 1}^{N_s}  (\boX_i - \boXbar)(\boX_i - \boXbar)^T.
 \label{S}
\end{equation}
This estimator is a matrix-variate random variable and in \citet{SH15}, we propagated the uncertainty on the true covariance $\boSig$ into parameter inference, by transferring the randomness of $\sfS$ onto $\boSig$ by virtue of a prior and Bayes' theorem. This lead to the likelihood of a $p$-dimensional dataset $\obsboX$, conditioned on the mean $\bomu$ and $\sfS$ from $N_s$ simulations: 
\begin{equation}
P(\obsboX | \bomu, \sfS, N_s) = 
\frac{\bar{c}_p |\sfS|^{-1/2}}{\left[ 1 + \frac{Q }{N_s-1}\right] ^{\frac{N_s}{2}}}.
  \label{cosmo_tdistrib}
\end{equation}
This is an ellipsoidally contoured $t$-distribution, using
\begin{equation}
Q = (\obsboX -\bomu)^T\sfS^{-1} (\obsboX - \bomu),
\end{equation}
and the normalization
\begin{equation}
 \bar{c}_p = \frac{\Gamma\left(  \frac{N_s}{2}   \right)}{ \left[\pi (N_s-1)\right]^{p/2} \Gamma \left( \frac{N_s-p}{2} \right) }.
 \label{cbar}
\end{equation}
Previous approximate results had been proposed in a series of papers \citep{Hartlap,Dodelson,TJK,TJ,Percival14} which employ only the first two moments instead of the entire distribution of $\sfS$, by construction maintaining a Gaussian likelihood.
In comparison to a Gaussian, a $t$-distribution has broader wings, and a more peaked core. 
Its width is primarily set by the number of simulations, as these determine how certain our estimate of the true covariance matrix is. The lower the number of simulations, the wider is the $t$-distribution which reflects the systematic loss of information due to estimating the covariance matrix. The obvious question in this framework is therefore how much information is lost for a certain finite $N_s$, in comparison to a known covariance matrix and this is the question we address in this paper.

We illustrate this loss with a toy model in Fig.~\ref{ToyModel} for a survey with similar characteristics to ESA's Euclid weak lensing survey \citep{Euclid}. Our agnostic model for Euclid assumes 10 redshift bins, leading to approximately $p = 1500$ data points. For Fig.~\ref{ToyModel}, we use $N_s = 1550$ simulations and $N_p = 50$ parameters that include cosmological parameters and a representative number of nuisance parameters. The relative uncertainties due to estimated covariance matrices do not depend on any further physical specifications. The solid contour is the joint $1\sigma$ credible region of two arbitrary parameters $\theta_1$ and $\theta_2$, derived from an assumed true covariance matrix. In contrast, the open contours are derived from different draws of estimated covariance matrices, and are systematically inflated, due to losing information when estimating the covariance matrix.

We will assess this loss by developing a Fisher matrix approximation to the $t$-distribution in order to completely include the uncertainty of $\boSig$. However, we will not stop at the Fisher matrix level, as previous works have done, as an important inequality enforces that the lost information of joint credibility regions and Figures of Merit (FoM) cannot be calculated from the information lost in a Fisher matrix: Jensen's inequality describes the fundamental problem that non-linear functions and averages of estimators do not commute. For any matrix variate function $f$ of the estimated covariance matrix $\sfS$, Jensen's inequality reads
\begin{equation}
 f( \langle\sfS \rangle) \leq  \langle f(\sfS) \rangle \\
 \label{Jensen}
 \end{equation}
for convex functions $f$, and the inequality inverts for concave functions. Here, angular brackets denote averaging, and the equality holds for linear functions only. This inequality implies that the mean loss of information for credible regions of jointly estimated parameters cannot be assessed by debiasing a Fisher matrix, because joint credibility regions are non-linear functions of the Fisher matrix. Likewise, the mean loss of information in the figure of merit cannot be predicted from the loss of information in the Fisher matrix, again due to the non-linear dependence between the two. We find it therefore necessary to extend the works of \citet{Dodelson,TJK,TJ}, which focused on Fisher matrices; here, we calculate the loss of information directly from the FoM and parameter credibility contours. We will find that the number of simulations needed to acquire a certain precision in the FoM or parameter credibility contours (or related quantities) depends on the estimator, but is generally lower than predicted by previous forecasts. This is mainly because we will use the more accurate $t$-distribution Eq.~(\ref{cosmo_tdistrib}) which was not available to previous forecasts.

We establish the mathematical framework in Sects.~ \ref{Fish_sect}-\ref{FoM_sect} and then validate it by a full Monte Carlo Markov Chain run in Sect.~\ref{Appl}. Conclusions for the information lost in current and future surveys are given in Sect.~\ref{Appl} and the discussion at the end of the paper.
 
\section{Fisher matrix}
\label{Fish_sect}

We begin by deriving the Fisher matrix of the $t$-distribution Eq.~(\ref{cosmo_tdistrib}). We assume that only the mean $\bomu$ depends on $N_p$ parameters $\theta_1\ldots \theta_{N_p}$, but the covariance matrix is parameter independent. This is strictly not the case, but is almost universally assumed, and is an excellent approximation as in typical cases the extra information from the parameter dependence of the covariance matrix is very small \citep{Tegmark:1996bz}. 

We employ Einstein's summation convention and let roman indices run over the data, $i \in [1,p]$ and greek indices over the parameters, $\alpha \in [1,N_p]$. The Fisher matrix is then
\begin{equation}
\sfF(\sfS)_{\alpha\beta} = \left\langle (\ln P)_{,\alpha}(\ln P)_{,\beta} \right\rangle
\end{equation}
where commas denote partial derivatives and the angular brackets denote the marginalization over the data by integrating over the $t$-distribution, with $\sfS$ kept fixed. Note that $\sfF(\sfS)$ is a random object, since $\sfS$ is. Evaluating the gradients of Eq.~(\ref{cosmo_tdistrib}), we have
\begin{equation}
\sfF_{\alpha\beta}(\sfS) = \left(\frac{N_s}{N_s-1}\right)^2\, \bar c_p  |S|^{-\half} S^{-1}_{ij} S^{-1}_{kl} \mu_{j,\alpha} \mu_{l,\beta} \,I_{ik},
\end{equation}
where we defined the integral
\begin{equation}
I_{ik}\equiv  \int \, \mathd^p X \, \frac{1}{\left(1+\frac{Q}{N_s-1}\right)^{N_s/2+2}}\,(\boX-\bomu)_i (\boX-\bomu)_k.
\label{Ikint}
\end{equation}
To solve the integral, we isotropize the data by diagonalising and whitening $\sfS$. Using an orthogonal rotation matrix $\sf{R}$, $\sfS^{-1}=\sf{R}^T \Lambda \sf{R}$, where $\sf{\Lambda}$ is diagonal.  Hence
\begin{equation}
Q =  z_j \Lambda_{jk}z_k
\end{equation}
where $z_i=R_{ij}(\boX-\bomu)_j$ and the Jacobian from $X_i$ to $z_i$ is unity. Suspending the summation convention, we whiten the covariance by rescaling $Z_i = \Lambda_{ii}^{\half}z_i$, leading to
\begin{equation}
I_{ik} = |S|^{\half}\,\sum_{st} \frac{R_{si}R_{tk}}{\Lambda_{tt}^{\half}\Lambda_{ss}^{\half}} J_{st},
\end{equation}
where the integral is
\begin{equation}
J_{st} = \int \, \mathd^p Z \frac{Z_s  Z_t}{\left[1+\frac{1}{N_s-1}\,\sum_{q} Z_q^2\right]^{N_s/2+2}}. 
\end{equation}
By symmetry, we now have $J_{st}=0$ unless $s=t$, in which case the integral depends only on $r^2 \equiv \sum_q Z_q^2$.
We therefore transform to polar coordinates,
\begin{equation}
\mathd^pZ = r^{p-1}\mathd r\sin^{p-2}\phi_1\,\sin^{p-3}\phi_2\ldots \sin\phi_{p-1}\mathd\phi_1\ldots \mathd\phi_p.
\end{equation}
The angular part of the integral is $A/p$, where 
\begin{equation}
A = \frac{2\pi^{p/2}}{\Gamma\left(\frac{p}{2}\right)},
\end{equation}
is the area of the unit sphere in the $p$-dimensional space of data points.
The integral over $r$ is
\begin{equation}
\int_0^\infty \frac{r^{p+1} \mathd r}{\left(1+\frac{r^2}{N_s-1}\right)^{\frac{N_s}{2}+2}} = \frac{\Gamma\left(\frac{N_s-p+2}{2}\right)\Gamma\left(\frac{p}{2}+1\right)}{2 (N_s-1)^{-(\frac{p}{2}+1)} \Gamma\left(\frac{N_s}{2}+2\right)}.
\end{equation}
The gamma functions simplify when combined with the normalization $\bar c_p$ from Eq.~(\ref{cbar}), giving
\begin{equation}
\sfF_{\alpha\beta}(\sfS) = \frac{N_s(N_s-p)}{(N_s-1)(N_s+2)}\,\sum_{ijkls} \frac{R_{si}R_{sk}}{\Lambda_{ss}} S^{-1}_{ij} S^{-1}_{kl} \mu_{j,\alpha} \mu_{l,\beta},
\end{equation}
where the sum over the $R$ and $\Lambda$ terms is $S_{ik}$. Introducing the $(p\times N_p)$ matrix $\sfA$
\begin{equation}
 \sfA = 
 \begin{pmatrix}
 \bomu,_1^T \\
 \vdots \\
 \bomu,_{N_p}^T \\
 \end{pmatrix}
\end{equation}
whose row vectors are the $N_p$ derivatives of the mean, the Fisher matrix of the $t$-distribution is therefore
\begin{equation}
\sfF(\sfS) = \frac{N_s(N_s-p)}{(N_s-1)(N_s+2)}\, \sfA \sfS^{-1}\sfA^T.
\label{FS}
\end{equation}
This Fisher matrix describes the average curvature of $t$-distribution around its peak, if different data sets $\boX$ are drawn. It keeps the covariance matrix $\sfS$ fixed and is an approximation to the width of the $t$-distribution. Therefore, it quantifies the lost information due to $\sfS$ being estimated from a finite number of simulations.

\begin{figure}
\includegraphics[width=0.45\textwidth]{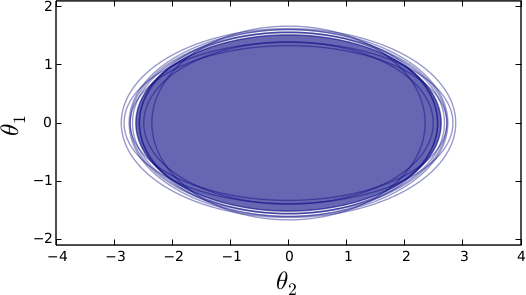} 
\caption{Scaling the confidence contours with Eq.~(\ref{scale_axes}) reduces their size to the optimal case of a known covariance matrix. The remaining scatter is then described by Eq.~(\ref{deb_area_var}) and depicted in Fig.~\ref{AreaBias} for a Euclid like survey.
}
\label{AreaCorrection}
\end{figure}

\begin{figure*}
\includegraphics[width=0.47\textwidth]{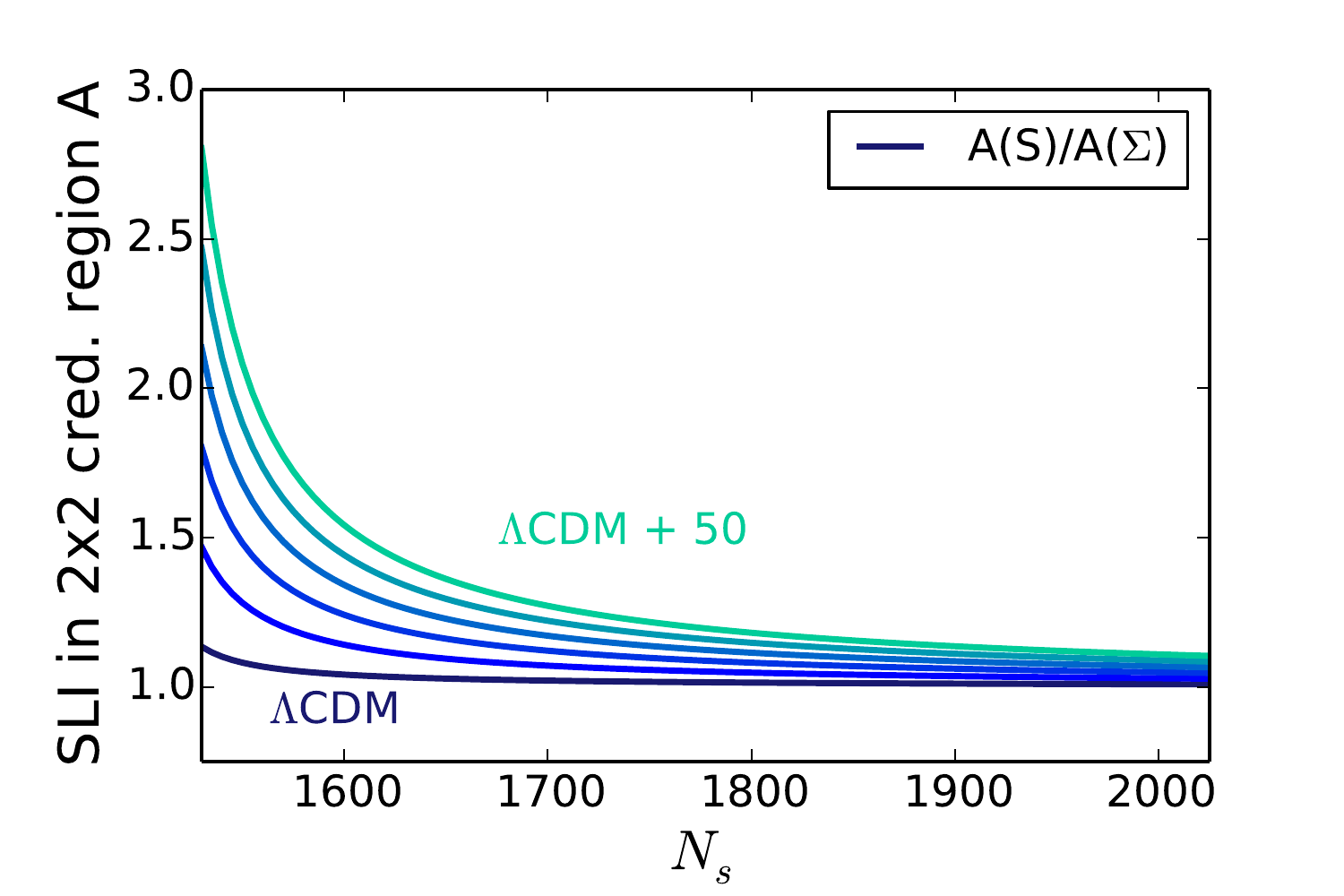} 
\includegraphics[width=0.47\textwidth]{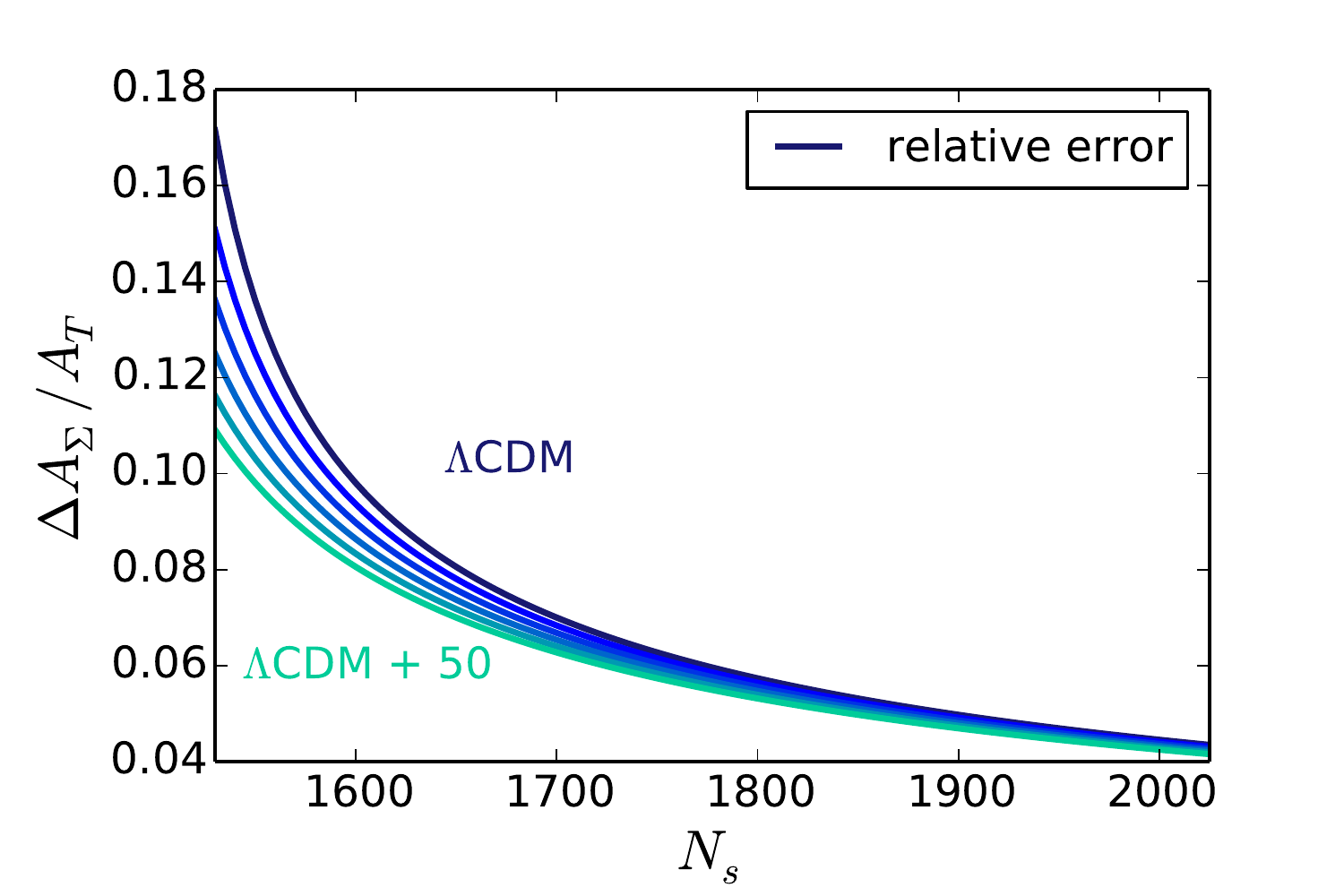} 
\caption{Left: systematically lost information (SLI) in each $2\times2$ parameter credibility contour, as e.g.\ displayed in a triangle plot, when a total of $N_p > 2$ parameters are estimated. The SLI describes the factor by which the credibility contours are systematically inflated, see also Fig.~\ref{ToyModel}. Right: the statistical uncertainty of forecasting the optimally achievable credibility contours, when extrapolating from an initial estimator of $N_s$ simulations via Eq.~(\ref{deb_area_var}). This produces the open contours of Fig.~\ref{AreaCorrection} and the quantity here plotted corresponds to their observed scatter in Fig.~\ref{AreaCorrection} around the true contour.
This plot assumes a Euclid-like number of data points, $p = 1500$. Starting at $N_p = 6$ for the $\Lambda$CDM line, the other lines subsequently add nuisance parameters in steps of 10.
}
\label{AreaBias}
\end{figure*}

A completely different use of this Fisher matrix would be to construct from it an estimator of the unknown \emph{true} Fisher matrix, $\sfF(\boSig) = \sfA \boSig^{-1} \sfA^T$ with the aim of undoing the lost information in the estimation process. In a real data analysis, such lost information can only be captured by increasing the number of simulations. One should therefore not attribute any other meaning to such a Fisher matrix than being the hypothetically optimal case that could be achieved for $N_s \rightarrow \infty$.  

In a Bayesian way, estimators of unknown quantities would ideally be constructed by marginalizing over their full distribution, conditional on all available information. In this paper, we will however soon deal with non-linear functions of estimators, where their full distribution becomes analytically intractable. Fortunately, a distribution can be written in terms of its moments, provided that they are finite. We therefore employ a moment-wise description of various estimators, as it is analytically tractable. By sampling the random processes under consideration, we furthermore find that working with the first two moments of these distributions already captures the essential information that we seek.

We now estimate the Fisher matrix for $\boSig$, given that we only know $\sfS$.
Since $\sfS^{-1}$ follows an inverse Wishart distribution, the distribution of $\sfF(\sfS)$ is also inverse Wishart, with \citep{GuptaNagar}
\begin{equation}
 \sfA\sfS^{-1}\sfA^T \sim \mathcal{W}^{-1}_{N_p}\left(n-p+N_p, (\sfA\boSig^{-1} n \sfA^T)\right).
\end{equation}
where $n = N_s-1$.
The mean Fisher matrix for a $t$-distribution, when also averaged over different samples of $\sfS$, is therefore
\begin{equation}
 \langle \sfF(\sfS)\rangle_{_\sfS} = \frac{N_s(N_s-p)}{(N_s-1)(N_s+2)} \langle \sfA\sfS^{-1}\sfA^T \rangle_{_\sfS} .
\end{equation}
where by property of the inverse Wishart distribution
\begin{equation}
 \langle \sfA\sfS^{-1}\sfA^T\rangle = \frac{n}{n-p-1} (\sfA\boSig^{-1}\sfA^T),
 \label{Sav_pref}
\end{equation}
where the number of data points $p$ appears because it is the dimension of the data covariance matrix $\boSig$. Our interest lies in estimating the true $\sfA\boSig^{-1}\sfA^T$. Consequently, we have
\begin{equation}
 \langle \sfF(\sfS) \rangle_\sfS = \frac{N_s(N_s-p)}{(N_s+2)(N_s-p-2)} (\sfA\boSig^{-1}\sfA^T).
 \label{Fmean}
\end{equation}
We therefore see that the Fisher matrix of the estimator $\sfS$ (random through its dependence on $\sfS$) is biased with respect to the true Fisher matrix $\sfA\boSig^{-1}\sfA^T$. This bias is however not an unexpected nuisance,  rather it describes the systematic loss of information on the parameters, due to estimating $\sfS$ from a finite number of simulations. We therefore often refer to the bias as systematically lost information (SLI) when emphasising its physical meaning. Removing this bias corresponds to a hypothetical restoration of the lost information, and a reduction of the bias can be achieved by increasing $N_s$. Eq.~(\ref{Fmean}) allows a cost-benefit analysis to determine whether this is worth doing.

We therefore divide the Fisher matrix Eq.~(\ref{FS}) by the prefactor of Eq.~(\ref{Fmean}) in order to mimic restoring the lost information, leading to 
\begin{equation}
 \sfF_\boSig(\sfS) = \frac{N_s-p-2}{N_s-1}\sfA\sfS^{-1}\sfA^T.
 \label{debF}
\end{equation}
We keep the subscript $\boSig$ to denote that this Fisher matrix is now an estimator that extrapolates from the known $\sfS$ to the Fisher matrix for the optimal case where $\boSig$ is known. The main difference between Eq.~(\ref{FS}) and Eq.~(\ref{debF}) is that $\sfF(\sfS)$ includes the SLI due to uncertainty of the estimated covariance matrix, i.e.\ it approximates the broad $t$-distribution, whereas  $\sfF_\boSig(\sfS)$ is constructed to \emph{restore} the SLI, because it is an estimator of the true Fisher matrix. $\sfF_\boSig(\sfS)$ will approximate the narrower Gaussian distribution of the parameters. Loosely speaking,  $\sfF(\sfS)$ approximates the outer open contours, and $\sfF_\boSig(\sfS)$ approximates the inner solid contour in Fig.~\ref{ToyModel}, where the caveat is that Eq.~(\ref{debF}) corresponds only to a `debiased' estimator of the true Fisher matrix itself, or any linear function of it. It therefore restores lost information only in the Fisher matrix itself. Unbiased parameter confidence contours, figures of merit and inverse Fisher matrices cannot be estimated from it, as these are non-linear functions of the Fisher matrix. Biases in these will reappear due to Jensen's inequality, even if $\sfF_\boSig(\sfS)$ itself is debiased. As the bias parametrizes the lost information in this context, we therefore proceed to calculate the SLI of these quantities explicitly.

\subsection{Inverse Fisher matrices}
The inverse of the Fisher matrix approximates the mean parameter covariance matrix, through the inverse of Eq.~(\ref{Fmean}),
\begin{equation}
 (\langle \sfF \rangle_\sfS)^{-1} = \frac{(N_s+2)(N_s-p-2)}{N_s(N_s-p)} (\sfA\boSig^{-1}\sfA^T)^{-1},
 \label{C_ES}
\end{equation}
whose bias is independent of the number of parameters $N_p$, as opposed to the parameter covariance matrix found by \citet{TJ}. Further, the bias factor of Eq.~(\ref{C_ES}) cannot diverge, as the Wishart distribution requires $N_s>p$ anyway. An estimator that restores the SLI for the inverse of the average peak curvature, is therefore
\begin{equation}
 \sfF^{-1}_{\boSig,{\rm peak}}(\sfS) = \frac{(N_s-1)}{(N_s-p-2)} (\sfA\sfS^{-1}\sfA^T)^{-1},
 \label{deb_peakcurv}
\end{equation}
which is indeed the inverse of the Fisher matrix Eq.~(\ref{debF}).

However, a second estimator for the true inverse Fisher matrix can be formed by taking the matrix inverse of the conditional Eq.~(\ref{FS}) directly,
\begin{equation}
\sfF^{-1}(\sfS) = \frac{(N_s-1)(N_s+2)}{N_s(N_s-p)} (\sfA \sfS^{-1}\sfA^T)^{-1}.
\label{FSinv}
\end{equation}
Again, this inverse Fisher matrix is random and follows a Wishart distribution \citep{GuptaNagar}:
\begin{equation}
 (\sfA\sfS^{-1}\sfA^T)^{-1} \sim \mathcal{W}_{N_p}\left(n-p+N_p, (\sfA\boSig^{-1} n \sfA^T)^{-1}\right).
 \label{distFinv}
\end{equation}
Its mean is therefore
\begin{equation}
\langle \sfF^{-1}(\sfS)\rangle = \frac{(N_s-p+N_p-1)(N_s+2)}{N_s(N_s-p)}  \,(\sfA \boSig^{-1}\sfA^T)^{-1},
\label{Cmarg}
\end{equation}
whose bias depends on the number of parameters, similar to the estimator found by \citet{TJ}. Its equivalent that restores SLI in a single estimated covariance matrix from Eq.~(\ref{FSinv}) is then
\begin{equation}
\sfF^{-1}_\boSig(\sfS) = \frac{(N_s-1)}{(N_s-p+N_p-1)}  \,(\sfA \sfS^{-1}\sfA^T)^{-1}.
\label{deb_Cmarg}
\end{equation}
The estimators Eq.~(\ref{C_ES}) and Eq.~(\ref{Cmarg}) differ numerically because averaging and inverting are mathematical operations that do not commute. Physically, the estimators Eq.~(\ref{C_ES}) and Eq.~(\ref{Cmarg}) correspond to different quantities, and it depends on the scientific context which of the two is of interest. Eq.~(\ref{C_ES}) is the inverse of the average curvature of the $t$-distribution around its peak, marginalized over data $\boX$ and over covariance estimators $\sfS$ which set the width of a $t$-distribution. Eq.~(\ref{Cmarg}) describes how the inverse Fisher matrix for a single $\sfS$ scatters if the parameter inference, conditional on an estimator $\sfS_1 \sim \mathcal{W}_p(n,\boSig)$ were repeated for other samples $\sfS_i \sim \mathcal{W}_p(n,\boSig)$. Eq.~(\ref{Cmarg}) then describes that if one calculates an inverse Fisher matrix for each sample $\sfS_i$ individually, the mean will be biased due to the lost information. As a side note, our inverse Fisher matrix Eq.~(\ref{Cmarg}) differs from the estimator given in \citet{TJ} because we work with the Bayesian $t$-distribution, whereas \citet{TJ} use a systematically inflated Gaussian likelihood following \citet{Hartlap}. Our main difference is however that we regard the Fisher matrix only as an intermediate step, needed to derive the quantities presented in the upcoming sections.

\section{SLI in parameter credible regions}

Typically, a data analysis has just a single estimated covariance matrix $\sfS$ at its disposal. Ultimately, our interest in it is to calculate parameter credible regions. For deriving the SLI of credible regions and the remaining scatter after hypothetically restoring it, the estimator Eq.~(\ref{FSinv}) for the inverse Fisher matrix is therefore the correct choice. Since the inverse Fisher matrix is random, its eigenvectors will scatter, leading to a reorientation of the credible regions that are derived from it. Additionally, the volume of credible regions will vary. Here we concentrate on the latter.

We employ a geometrical interpretation of covariance matrices. The Fisher matrix approximation is an ellipsoidally-contoured likelihood with 
\begin{equation}
 L(\both) \approx \mathcal{C} \exp\left[-\half (\both-\hat\both)^T \sfF(\sfS) (\both-\hat\both)\right],
\end{equation}
where $\mathcal{C}$ is a normalization constant and $\hat\both$ are the best-fitting parameters. The quadratic form 
\begin{equation}
 (\both-\hat\both)^T \sfF(\sfS) (\both-\hat\both) = \chi_r^2(\alpha),
\end{equation}
describes hyper-ellipsoids, centered on $\hat\both$, encasing $1-\alpha$ of the total probability. Their axes are $\boldsymbol{a}_i^\pm = \pm \sqrt{\chi_r^2(\alpha)\lambda_i}\boldsymbol{e}_i$, where $\lambda_i$ is the $i$th eigenvalue of $\sfF^{-1}$, and $\boldsymbol{e}_i$ is the corresponding unit-length eigenvector. The degrees of freedom $r$ of the $\chi^2$-distribution are the number of parameters to be estimated. 

The volume of a $r$-dimensional unit-sphere is
\begin{equation}
 V(r) = \frac{\pi^{r/2}}{\Gamma(\frac{r}{2} +1 )}.
\end{equation}
The inverse Fisher matrix then maps this unit hyper-sphere onto the hyper-ellipsoids that describe the parameter confidence levels. The volume of the mapped and distorted hyper-sphere then changes by the determinant of the inverse Fisher matrix, such that in $r$ dimensions, the volume of an hyper-ellipsoid of parameter credible regions is
\begin{equation}
 V_e(r) = \frac{\pi^{r/2}}{\Gamma(\frac{r}{2} +1 )} \left[ \chi^2_r(\alpha) \right]^{\frac{r}{2}} \sqrt{|\sfF^{-1}_{r\times r}|} .
 \label{E_vol}
\end{equation}
If we want to measure $N_p$ parameters then $|\sfF^{-1}_{N_p \times N_p}|$ is needed to describe the total parameter volume, whereas if we are only interested in an $r\times r$ submatrix of $\sfF^{-1}_{N_p\times N_p}$, then using its determinant in Eq.~(\ref{E_vol}) will describe the parameter volume of the remaining $r$ parameters, where the other parameters have been marginalized, due to discarding rows and columns from the inverse Fisher matrix. Consequently, for assessing the SLI of joint credible regions, we require the bias and variance of submatrices of $\sfF^{-1}$, and of their determinants. 

\begin{figure*}
\includegraphics[width=0.47\textwidth]{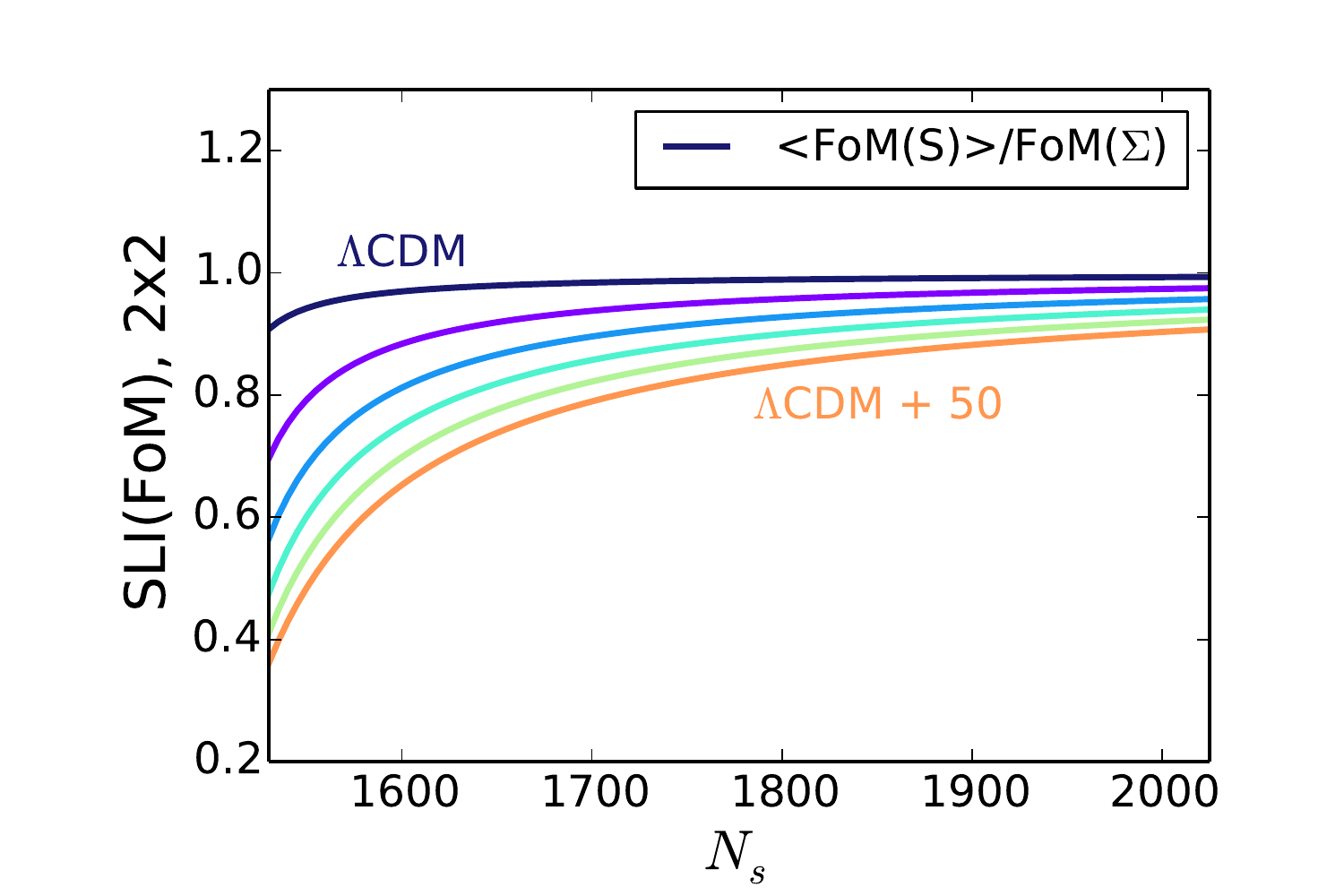} 
\includegraphics[width=0.47\textwidth]{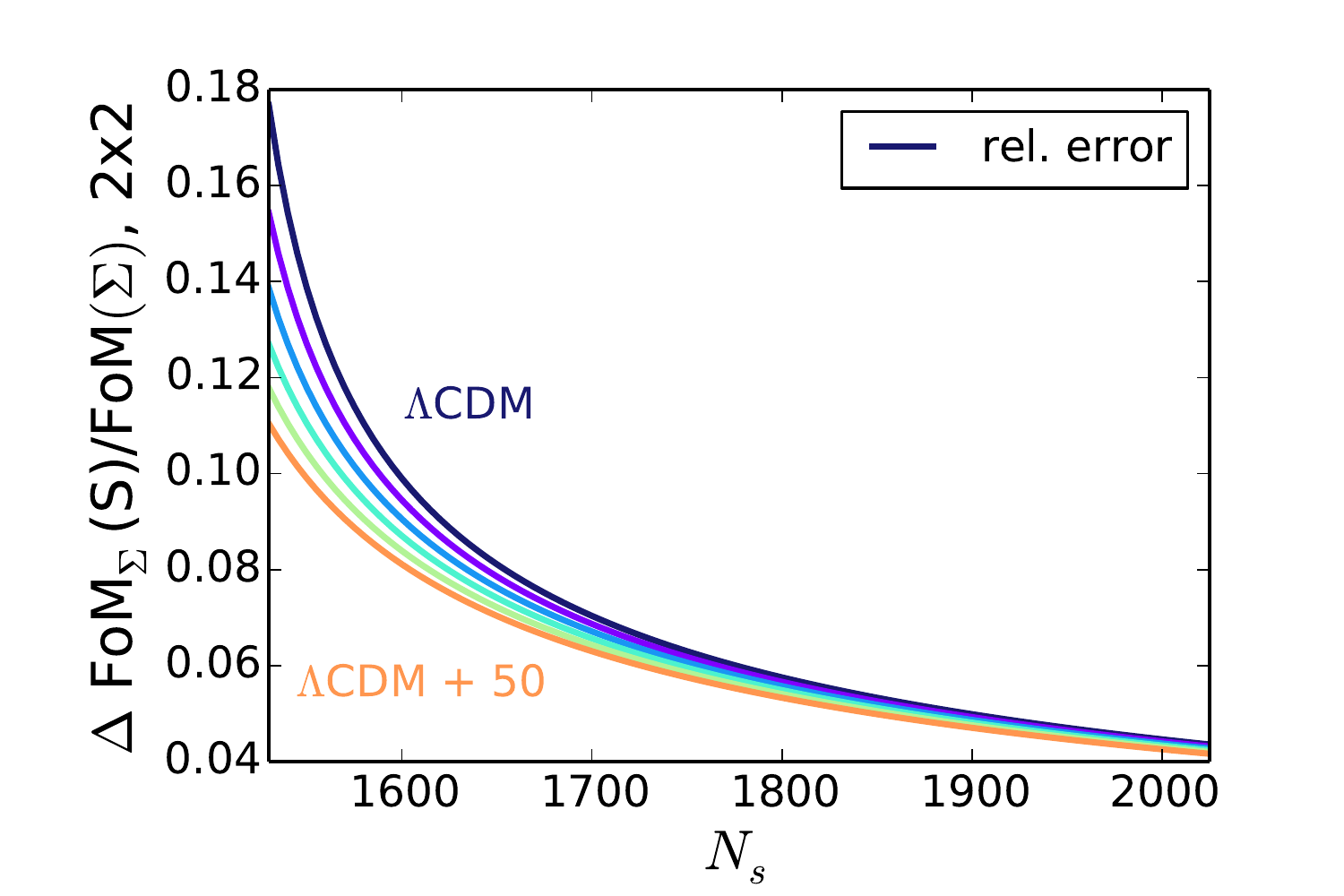} 
\caption{Same Euclid-like setup as in Fig.~\ref{AreaBias}, but now analysing the Figure of Merit (FoM) of 2 parameters, if a total of $N_p$ parameters as indicated by the linestyle are estimated. Left: The SLI decreases if more simulations are run, meaning the figure of merit increases towards its optimal value (which is unity in the plot). Hypothetically restoring the SLI is achieved with Eq.~(\ref{deb_fom}) and the right panel depicts the relative error Eq.~(\ref{relfom}) which describes how accurately the true FoM can be predicted when starting from a an estimator with $N_s$ simulations.
}
\label{FoMbias}
\end{figure*}

Fortunately, $\sfF^{-1}(\sfS)$ is Wishart distributed, and $r \times r$ submatrices from the diagonal of a Wishart matrix are again Wishart distributed with the same degrees of freedom but dimension $r$ \citep{AndersonTW}. Defining the pre-factor of Eq.~(\ref{FS}) as $\lambda = N_s(N_s-p)/[(N_s-1)(N_s+2)]$, we therefore have that an $r \times r$ submatrix of $\sfF^{-1}$ is distributed according to
\begin{equation}
\left( \lambda \sfF^{-1}(\sfS) \right)_{r\times r} \sim \mathcal{W}_r(n-p+N_p, (\sfA \boSig^{-1}n\sfA^T)^{-1}_{r\times r}).
\end{equation}
Further, the $h$th moments of the distribution of a Wishart matrix's determinant are known to be \citep{GuptaNagar}
\begin{equation}
 \langle |\sfF^{-1}_{r\times r}(\sfS)|^h\rangle = |(\sfA\boSig^{-1}\sfA^T)^{-1}_{r\times r}|^h B_r(h),
 \label{dets}
\end{equation}
where the Fisher matrix derived from the true covariance matrix is $\sfF(\boSig) = \sfA\boSig^{-1}\sfA^T$ and we defined the functions
\begin{equation}
 B_r(h) \equiv \frac{2^{rh}}{\lambda^{rh} (N_s-1)^{rh}}\frac{\Gamma_r\left( \frac{\nu}{2} + h\right)}{\Gamma_r\left(\frac{\nu}{2}\right)},
 \label{B}
\end{equation}
with $\nu = N_s-1-p+N_p$.
Putting $h = 1/2$ in Eq.~(\ref{dets}) yields the SLI of the mean of $\sqrt{|\sfF^{-1}_{r\times r}(\sfS)|}$ with respect to a known covariance matrix
\begin{equation}
 \mathrm{SLI}_\mathrm{hypervol.} = B_r(\half).
 \label{SLIhyp}
\end{equation}
In other words, Eq.~(\ref{SLIhyp}) is the systematically lost information of the $r$-dimensional hypervolume within ellipsoids of constant credibility. Since the SLI is positive, the parameter volume that is compatible with the data, is inflated, compared to the case of a known covariance matrix.
The estimator for the parameter volume as allowed by the true covariance matrix $\boSig$ is therefore
\begin{equation}
 V_\boSig(r,\sfS) = \frac{V_e(r,S)}{B_r\left(\half \right)},
 \label{deb_Vol}
\end{equation}
where $V_e(r,S)$ is defined by Eq.~(\ref{E_vol}) using $\sfF^{-1}(\sfS)$ from Eq.~(\ref{FSinv}). The variance of this debiased parameter volume is then
\begin{equation}
 \var(V_\Sigma) = |\sfF^{-1}_{r \times r}(\boSig)| \frac{ [\pi\cdot\chi^2_r(\alpha)]^r }{[\Gamma(\frac{r}{2}+1)]^2} \left[ \frac{B_r(1)}{B^2_r(\half)} - 1\right].
 \label{deb_area_var}
\end{equation}
Specializing to two dimensions, $r = 2$, yields the area of credible regions for parameter pairs as displayed in a triangle plot. Eq.~(\ref{E_vol}) is then the area of an ellipse $A = \pi |\boldsymbol{a}_1| |\boldsymbol{a}_2|$ where the the semi-axes are given by
\begin{equation}
 \boldsymbol{a}_i  = \sqrt{\lambda_i \chi^2_2(\alpha)} \, \boldsymbol{e}_i,
\end{equation}
We can restore the SLI as in Eq.~(\ref{deb_Vol}) for an ellipse by rescaling its semi-axes:
\begin{equation}
 \boldsymbol{a}_i \rightarrow \frac{\boldsymbol{a}_i}{ \sqrt{B_2(\half)}  }.
 \label{scale_axes}
\end{equation}
We caution that this rescaling cannot be applied to parameter credible regions of a real dataset. It merely allows a forecast of how much information could be gained, if the covariance matrix were known.

Since $A_T = \pi\chi^2_2(\alpha) \sqrt{|\sfF^{-1}_{2\times 2}(\boSig)|}$ is the area encased by the true credible region, the statistical error of its estimator $A_\boSig$, relative to $A_T$ is
\begin{equation}
 \frac{\sqrt{\var(A_\Sigma)}}{\pi\chi^2_2(\alpha) \sqrt{|\sfF^{-1}_{2\times 2}(\boSig)|}} = \sqrt{ \frac{B_2(1)}{B^2_2(\half)} - 1}.
\end{equation}
The SLI and the relative error for ellipses is given by Eq.~(\ref{SLIhyp}) for $r=2$. This SLI is lost in each pair of jointly estimated parameters and is plotted in Fig.~\ref{AreaBias} for a Euclid-like survey. The left panel demonstrates that the systematically lost information for the credible regions compared to the ideal case of a known covariance matrix. We observe at first a steep decrease of the lost information when more simulations are run, meaning that running only a few more simulations than the minimally possible number, very quickly improves the constraining power. Afterwards, the curves flatten and an asymptotic domain is reached, where the remaining lost information is restored ever more slowly. In contrast, the number of nuisance parameters affects critically how many simulations are needed in order to keep the SLI per parameter-pair below a chosen limit. As the total number of free parameters increases, the SLI increases as well, and a reduction of nuisance parameters can be a more efficient way to reduce the impact of noise in estimated covariance matrices. Fig.~\ref{AreaBias} shows what could be gained if only more simulations were run, and this analysis will be useful in determining a sweetspot between nuiance parameters and number of simulations when planning future surveys. The right panel depicts the result of restoring the SLI in a forecasting experiment via the estimator $A_\boSig$. This estimator will scatter as depicted. However, its relative error decreases with the number of nuisance parameters as their uncertainty already inflates the size of the credibility regions.
Finally, for $r = 1, h = 1$, it follows from Eq.~(\ref{B}) that the SLI in the variance of each single parameter $i$ marginalized over all other parameters is
\begin{equation}
 \frac{\langle \sigma_i^2(\sfS) \rangle}{\sigma_i^2(\boSig)} = \frac{(N_s-1-p+N_p)(N_s+2)}{N_s(N_s-p)},
 \label{var}
\end{equation}
illustrating that for a given survey with fixed number of simulations $N_s$ and data points $p$, the lost information scales only linearly with the number of nuisance parameters.
\section{SLI in the Figure of Merit}
\label{FoM_sect}

The inverse of the parameter volume that is compatible with the data, is a useful figure of merit. In the Gaussian approximation, this volume is proportional to the square root of the determinant of the inverse Fisher matrix, where the inverse is taken, because ignoring rows and columns of it allows to quickly marginalize parameters. We therefore work with the figure of merit
\begin{equation}
 \fom = \frac{1}{\sqrt{|\sfF^{-1}_{r\times r}(\sfS)|}}.
 \label{FoM}
\end{equation}
Eq.~(\ref{FoM}) is the FoM conditional on $\sfS$, and has lost information due to the finite number of simulations in $\sfS$. The mean lost information with respect to the case of a known $\boSig$ follows from putting $h = -1/2$ in Eq.~(\ref{dets}). Dividing by the determinant of the true Fisher matrix yields the following estimator for the figure of merit that restores the SLI
\begin{equation}
 \fom_\boSig(\sfS) = \frac{1}{\sqrt{|\sfF^{-1}_{r\times r}(\sfS)|}} \frac{1}{B_r(-\half)}.
 \label{deb_fom}
\end{equation}

If $r = N_p$, Eq.~(\ref{FoM}) describes the figure of merit for the entire parameter set, if $r = 2$, Eq.~(\ref{FoM}) allows to define a figure of merit e.g.\ for the $w_0,w_a$ subspace of a generic dark energy model.
The variance of the estimated figure of merit with the restored SLI is then
\begin{equation}
 \var[\fom_\boSig(\sfS)] = \left[\fom(\boSig)\right]^2 \left[ \frac{B_r(-1)}{B_r^2(-\half)} -1 \right] ,
\end{equation} 
resulting in the relative error with respect to the true figure of merit
\begin{equation}
 \frac{\sqrt{ \var[\fom_\boSig(\sfS)] }}{\fom(\boSig)} = \sqrt{\frac{B_r(-1)}{B_r^2(-\half)} -1}.
 \label{relfom}
\end{equation}
The SLI and the remaining relative error after restoring the SLI are depicted in Fig.~\ref{FoMbias}.

\section{Application to current surveys}
\label{Appl}

\begin{figure}
\includegraphics[width=0.47\textwidth]{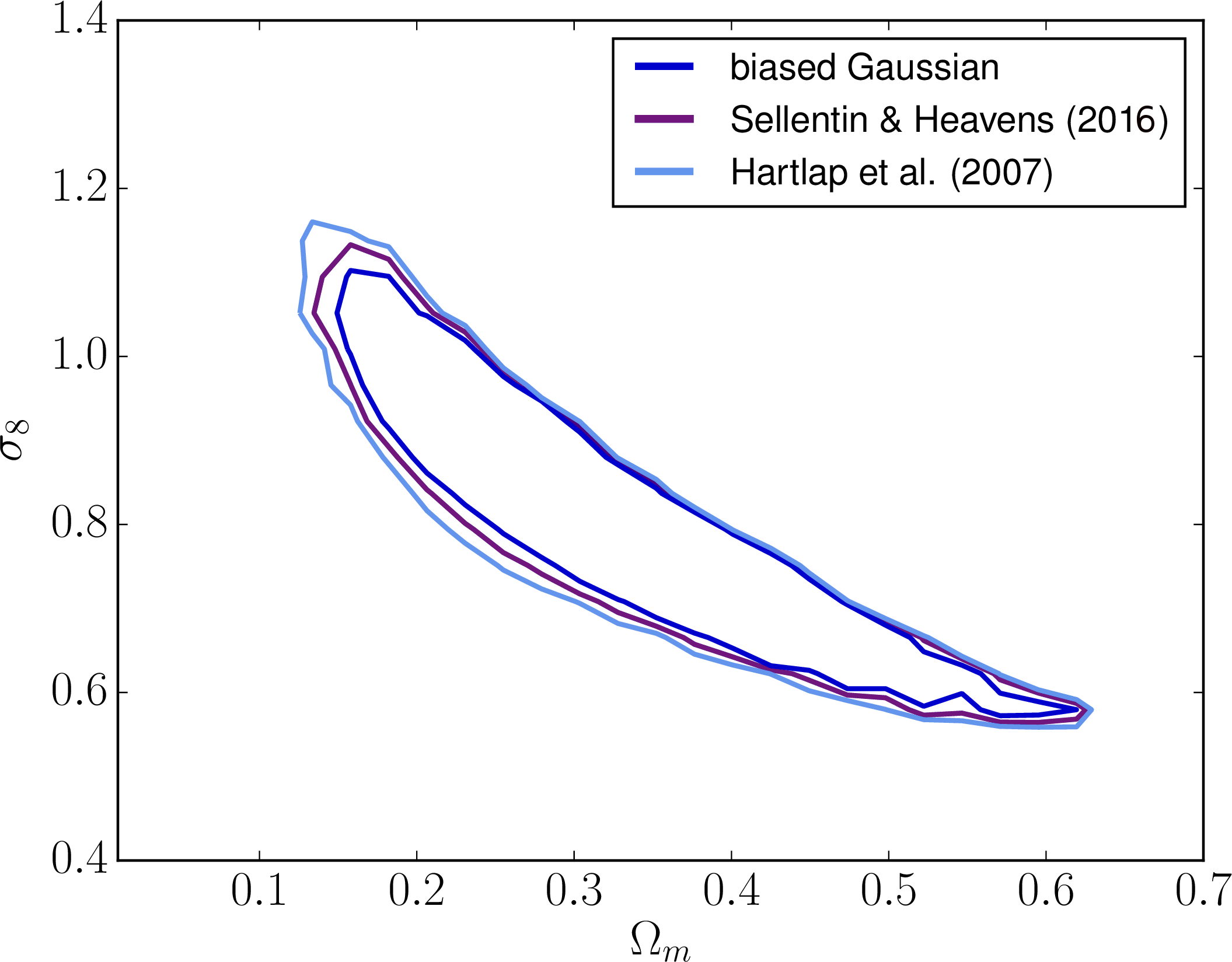} 
\caption{68\% credible regions for the matter density parameter $\Omega_m$ and the amplitude of the linear perturbations $\sigma_8$, derived from the Hartlap et al. likelihood (outer), from the Sellentin-Heavens likelihood (middle), and from the ideal Gaussian likelihood (inner), which deliberately ignores uncertainties in the estimated covariance matrix, in order to illustrate that the dominant fraction of the total parameter uncertainty arises from other effects than covariance uncertainty.}
\label{SHvHartlap}
\end{figure}

\subsection{KiDS and DES}
The KiDS-450 survey \citep{KiDS} employs $p = 130$ data points, and a covariance matrix estimated from 930 simulations. Estimating about 10 cosmological parameters, the parameter credibility contours are inflated at the insignificant level of 1\% or less. The statistical deviations of the contours from their true position is then about 3\%, allowing to infer that KiDS does not suffer from noise in the covariance matrix.

The DES science verification (SV) weak lensing analysis uses $p = 36$ data points and a covariance matrix estimated from $126$ simulations. For estimating about 10 cosmological parameters, we therefore detect a 10\% SLI in the DES 2D marginal contours, together with an additional 10\% statistical scatter. DES would need to increase the number of simulations to $\approx 450$ in order to reduce the lost information to $5\%$, keeping the dimension of the data vector fixed to $p = 36$. For the analysis of the full DES data set, the number of simulations needed can be estimated with Eq.~(\ref{deb_Vol}).

Since all of the previous results hold in the Fisher matrix approximation, we now check on their validity by running a full likelihood, comparing with our expectations. We choose the weak lensing measurement of the DES SV data, varying all parameters of $\Lambda$CDM together with two intrinsic alignment parameters. Physically, our setup is exactly that presented in \citet{DES}, and we modify the statistical analysis only. Fig.~\ref{SHvHartlap} depicts our results. The total uncertainty on the combination $\sigma_8-\Omega_m$ includes the degeneracy of the parameters in the $\Lambda$CDM model, the finite number of data points and their noise, and finally the uncertainty of the covariance matrix. Deliberately ignoring all statistical uncertainty in the covariance matrix, allows us to estimate how much of the error bars on $\sigma_8$ and $\Omega_m$ is due to the parameter degeneracies and the finiteness of the noisy data set. This produces the innermost blue contour seen in Fig.~\ref{SHvHartlap}, which gives a good estimate of the optimally achievable parameter bounds. Including the uncertainty of $\sfS$ by using the $t$-distribution produces the middle contour in Fig.~\ref{SHvHartlap}. For comparison, we also plot the contour derived with the procedure proposed by \cite{Hartlap}, which produces the outermost contour. The closeness of all contours in Fig.~\ref{SHvHartlap}, reveals that the primary source of parameter uncertainty is not the uncertainty of the covariance matrix, but rather the degeneracy of the parameters and the finiteness of the data set. This finding is in line with our findings in Fig.~\ref{AreaBias} and Fig.~\ref{FoMbias} where we found for Euclid that the number of nuisance parameters is the more pressing issue than the number of simulations for the covariance matrix. We furthermore see that our Fisher matrix predictions capture the trend of the credibility contours well: Updating from Hartlap et al.'s method to the $t$-distribution moves the contours in. Note that although the corrections seem small, the tightening of the credible regions is roughly equivalent to observing half a million (corresponding to 20\%) more galaxies. We made sure none of these results are due to numerical issues with running Monte-Carlo Markov Chains: multiple chains were run, which fulfil the Gelman-Rubin convergence test.  The exchange of the statistical analysis was achieved by reweighting the chains, instead of resampling, such that all differences in the contours come from the different statistical likelihoods, rather than from different sampling.

\begin{figure*}
\includegraphics[width=\textwidth]{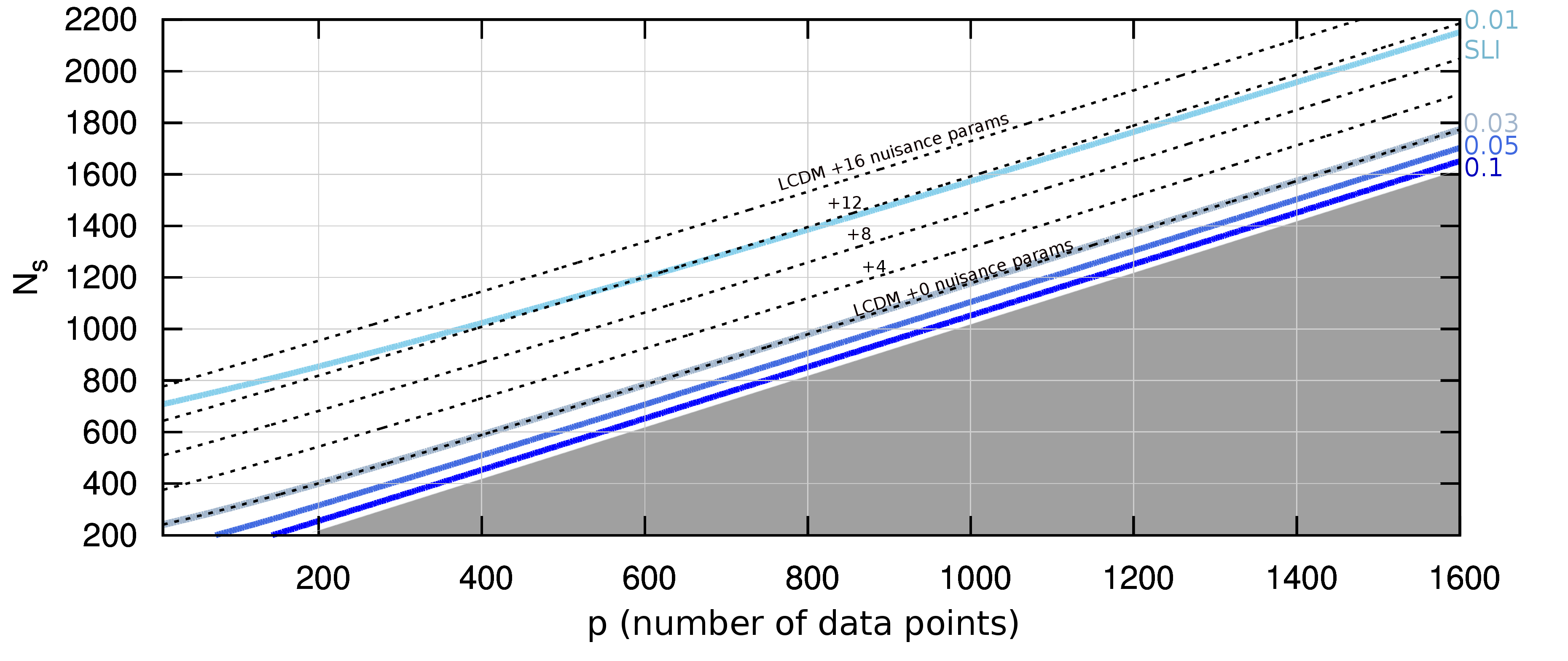} 
\caption{Orientation for future surveys: the number of simulations $N_s$ needed to reach a certain precision, given the number of data points $p$ and the number of parameters. The grey triangle is excluded, because at least as many simulations as data points have to be run. Its edge is the minimal number of simulations that need to be run. The thick blue lines mark a 6-parameter $\Lambda$CDM model without any nuisance parameters, labeled in blue hues by the different percentages of systematically lost information. The dashed black lines all refer to a 3\% loss of information, with an additional number of nuisance parameters as indicated in black. The quantity here plotted is the lost information in the variance of single parameters marginalized over all other jointly estimated parameters, see Eq.~(\ref{var}). The needed $N_s$ for reaching a comparable precision in joint parameter confidence contours, or in model comparison, is then slightly higher.
}
\label{comp}
\end{figure*}

\subsubsection{Flat vs non-flat Universe}

Finally, we note that one might expect the broader tails of the \cite{SH15} likelihood to have a significant effect on Bayesian evidence calculations.  We have tested this with the DES SV data, using nested sampling \citep{Multinest} within cosmosis \citep{cosmosis} to compute the Bayes factor for flat and non-flat Universes, with a prior range $\pm 0.2$ for the curvature parameter $\Omega_k$.   Both the \cite{SH15} and \cite{Hartlap} likelihood formally favour a non-flat Universe ($\ln(B)=0.17\pm0.09$ and $0.30\pm 0.07$ respectively), but neither significantly, and the difference in Bayes factors is insignificant in this case, but the sign of the change is as expected.

In the context of model comparison, we note that the $r$-dimensional Figure of Merit (and its associated parameter volume $V_r$) calculated earlier plays an interesting role, closely related to Lindley's paradox.  Consider two nested models, in which the simpler model A corresponds to a special point in B, setting $r$ parameters in B to specific values.    If the data have low probability in A, model A may still be preferred in a Bayesian analysis,  provided that the prior probability that the parameters are close to the special point is even smaller in B.
The Bayes factor essentially compares these two potentially small probabilities.  It is roughly the ratio of $\exp(-\chi^2/2)$ in model A to the probability in B that the special point lies within the volume $V_r$ when the centre of $V_r$ is chosen randomly from the prior volume.   For a uniform prior, this latter probability is essentially $V_r$ divided by the prior volume.

\section{Conclusions}
\label{dis}
If the data covariance matrix in parameter inference is estimated from $N_s$ simulations, its uncertainty leads to a loss of information on cosmological parameters. In this paper, we have quantified the systematically lost information (SLI) with a Fisher matrix approach starting from the $t$-distribution Eq.~(\ref{cosmo_tdistrib}) found in \citet{SH15}, which is the Bayesian solution to propagating the covariance matrix uncertainty into the parameters.

Within our approximations, the SLI corresponds to a `bias' of the $t$-distributions credible regions with respect to the narrower credible regions derived from a true covariance matrix. This lost information can be reduced in a real data analysis by increasing the number of simulations.  Due to Jensen's inequality, this SLI in the Figure of Merit and joint parameter credibility regions cannot be assessed by removing a bias in the Fisher matrix or its inverse, which is the level at which previous works have stopped \citep{Dodelson, TJ}. We therefore computed the SLI of these quantities directly, see Sects. \ref{FoM_sect} -\ref{Appl}. This allows us to estimate how far away a given survey is from its optimally achievable error bounds.

We summarize our scientific conclusions as follows. 
We find that for current weak lensing surveys, as KiDS-450 and DES SV, the parameter degeneracies and the finite data set are the dominating drivers of the uncertainties of cosmological parameters; uncertainty due to the covariance matrix is subdominant. KiDS systematically lost about 1\% of information, the analysis of the DES SV about 10\%. The exact numbers depend on the quantity for which the information loss is estimated, e.g.\ the variance of a single parameter, the credibility region of all jointly estimated parameter pairs, the entire parameter volume, or the Figure of Merit.

Concerning a Euclid-like weak lensing survey with $p = 1500$ data points, Eq.~(\ref{deb_Vol}) predicts that in a pure $\Lambda$CDM measurement Euclid would systematically lose only $1\%$ of information  for 1900 simulations, as compared to knowing the covariance matrix. This is only about twice as many simulations as have been run for KiDS.  An estimation of the optimally achievable error bound can then be achieved with a $4\%$ uncertainty. In the Bayesian framework of evaluating the $t$-distribution, this 4\% uncertainty does not need to be added to the parameter uncertainties anymore -- the $t$-distribution already accounts for this. The uncertainty merely refers to the reliability of predicting the optimally achievable error bound. For 10 additional nuisance parameters 2900 simulations are sufficient to only lose 1\% of systematic uncertainty, leading to a remaining scatter of 2\%.

In general, the Wishart nature of estimated covariance matrices already enforces that at least as many simulations as the dimension of the data set are run. We found for all estimators here considered, that a minor increase of the simulations over this minimally needed number already limits the lost information to below 10\% or 5\%. Afterwards, an asymptotic domain is reached, where a further increase of $N_s$ leads to an ever slower restoration of the remaining lost information. Fortunately, reducing the number of nuisance parameters reduces the lost information due to covariance matrix estimation much more quickly. In the case of fully marginalized variances of individual parameters, the SLI reduces linearly with a reduction of nuisance parameters. For all weak lensing surveys, this highlights the desirability of obtaining a better physical understanding of intrinsic alignments and shape measurement challenges, so that only a limited number of nuisance parameters are required.

For future surveys, we compress the findings in this paper into Fig.~\ref{comp}, which displays the needed number of simulations to reach a certain precision, given a targeted number of data points, and a known number of cosmological and nuisance parameters, as indicated by the line style.

In comparison to previous forecasts \citep{Dodelson,TJK,TJ,Percival14}, we find here significantly lower figures for the needed number of simulations in order to aquire a certain precision in parameter inference. This is mainly due to our use of the $t$-distribution, Eq.~(\ref{cosmo_tdistrib}) which is more accurate than the scaled Gaussian likelihood as proposed by \citep{Hartlap} and employed in \citet{Dodelson,TJK,TJ,Percival14}.

As a side effect, this study has produced all quantities needed for forecasts with estimated covariance matrices. Eq.~(\ref{FS}) gives the Fisher matrix, as e.g. needed for Euclid. 
 
The optimally achievable error bounds can then be obtained using the rescaling of ellipse-axes as in Eq.~(\ref{scale_axes}).

\section{Acknowledgements}
We thank Joe Zuntz, Sarah Bridle and Andrew Jaffe for discussions, support and comments. cosmosis can be found at https://bitbucket.org/joezuntz/cosmosis/wiki/Home.  This work was supported by a postdoc fellowship of the German Academic Exchange Service (DAAD), awarded to ES.

\bibliographystyle{mn2e}
\bibliography{TDist}

\label{lastpage} 
\bsp 
\end{document}